\documentclass[lettersize, journal]{IEEEtran}
\usepackage{amsmath,amsfonts}
\usepackage{algorithmic}
\usepackage{array}
\usepackage[caption=false,font=normalsize,labelfont=sf,textfont=sf]{subfig}
\usepackage{textcomp}
\usepackage{stfloats}
\usepackage{url}
\usepackage{verbatim}
\usepackage{hyperref}
\usepackage{graphicx}
\usepackage{booktabs}
\usepackage{multirow}
\usepackage{enumitem}
\DeclareMathOperator{\sgn}{sgn}
\def\BibTeX{{\rm B\kern-.05em{\sc i\kern-.025em b}\kern-.08em
    T\kern-.1667em\lower.7ex\hbox{E}\kern-.125emX}}
\usepackage{balance}
\newcolumntype{L}{>{$}l<{$}}
\newcolumntype{C}{>{$}c<{$}}
\newcolumntype{R}{>{$}r<{$}}

\def\eg{\textit{e.g.}}

\def\etc{\textit{etc}}

\begin{document}
\title{Focus on \textit{Change}: Mood Prediction by Learning Emotion Changes via Spatio-Temporal Attention}

\author{Soujanya~Narayana,~\IEEEmembership{Student~Member,~IEEE},~Ramanathan~Subramanian,~\IEEEmembership{Senior~Member,~IEEE},~Ibrahim~Radwan,~\IEEEmembership{Member,~IEEE},~and~Roland~Goecke,\IEEEmembership{~Senior~Member,~IEEE}\thanks{Soujanya Narayana, Ramanathan Subramanian, Ibrahim Radwan, and Roland Goecke are with the Human-Centred Technology Research Centre, Faculty of Science and Technology, Uni. Canberra, Bruce, ACT, Australia.}}


\markboth{Pre-print version}%
{Shell \MakeLowercase{\emph{et al.}}: Bare Demo of IEEEtran.cls for Signal Processing Society Society Journals}

\maketitle

\begin{abstract}
While \textit{emotion} and \textit{mood} interchangeably used, they differ in terms of duration, intensity and attributes. Even as multiple psychology studies examine the mood-emotion relationship, \textit{mood prediction} has barely been studied. Recent machine learning advances such as the \textit{attention} mechanism to focus on salient parts of the input data, have only been applied to infer emotions rather than mood. We perform mood prediction by incorporating both \textit{mood} and \textit{emotion change} information. We additionally explore spatial and temporal attention, and  parallel/sequential arrangements of the spatial and temporal attention modules to improve mood prediction performance. To examine generalizability of the proposed method, we evaluate models trained on the AFEW dataset with EMMA. Experiments reveal that (a) emotion change information is inherently beneficial to mood prediction, and (b) prediction performance improves with the integration of sequential and parallel spatial-temporal attention modules.
\end{abstract}


\section{Introduction}
Psychology and cognitive science studies reveal that emotions play an essential role in rational decision making, perception, and other cognitive processes~\cite{picard2000affective}. There is also a growing body of evidence that a healthy balance of emotions is integral to human intelligence and problem solving~\cite{tyng2017influences}. Although emotions have been studied by researchers for long, no single definition has been agreed upon. Emotions are considered as neurophysiological changes caused by external stimuli, and associated with feelings and behavioural changes~\cite{damasio1998emotion}. \textit{Mood} is another affective state, which is often considered synonymous with \textit{emotion}. Although there are no absolute boundaries between these two affective states, they differ in terms of duration, intensity and attributes. While \emph{emotion} is a short-term affective state lasting a few seconds or minutes, \emph{mood} denotes a long-term affective state lasting for minutes, hours or even days~\cite{jenkins1998human}. Mood akin to emotion is known to influence thought process, creativity and judgement \cite{picard2000affective}. 

While much research has been devoted towards emotion inference\footnote{We use the term \textit{inferring} instead of \textit{recognising} emotions/mood.} \cite{dzedzickis2020human, cowie2001emotion}, very few computational studies attempt mood prediction~\cite{katsimerou2015predicting}. The psychology literature (a) recognizes mood and emotion as distinct mechanisms that repeatedly trigger the arousal of each other~\cite{wong2016mood}, and (b) ascribes mood as a higher-level disposition that activates lower-level states such as emotions and beliefs~\cite{morris1992functional}. Nevertheless, joint modelling of mood and emotion for affect prediction has been neglected. 

From an affective computing perspective, \textit{mood inference} entails learning from data rich in affective state annotations, and assigning them to mood categories. The limited number of mood prediction studies can be attributed to the few databases with mood annotations. Popular affective databases such as HUMAINE~\cite{douglas2007humaine}, SEMAINE~\cite{mckeown2011semaine} and DREAMER~\cite{katsigiannis2017dreamer} are rich in emotion annotations, whereas barely any dataset exists with mood annotations-- these annotations denote the mood perceived by the human annotator rather than the actual mood expressed by the actor~\cite{katsimerou2016crowdsourcing}. Another reason for the limited availability of mood-labeled data is that affective labeling is onerous. While a large body of work leverages machine learning for emotion inference via  visual, acoustic, and textual data~\cite{ko2018brief, khalil2019speech, yoon2018multimodal}, very few explore mood inference.

Recent computer vision studies are considerably inspired by the human ability to naturally focus on \textit{salient} regions in complex scenes, known as \textit{attention}. Whilst developed for machine translation tasks~\cite{bahdanau2014neural}, attention is popularly incorporated in vision systems to improve their recognition performance~\cite{pmlr-v37-xuc15,hu2018squeeze}. The attention mechanism guides models to learn \emph{what} information is meaningful and \emph{where} to look for the same for a certain task. Attention has also been applied for emotion inference, and to guide models to focus on emotionally relevant parts of speech~\cite{ramet2018context}, video frames~\cite{ghaleb2020multimodal}, \etc. While attention has been applied for detecting mood disorders~\cite{huang2019attention}, it has not been employed for mood prediction. Mood inference is critical for applications such as diagnosis of mood-related disorders, mood regulation, and developing affective interfaces.

\begin{figure*}[!ht]
    \centering
    \includegraphics[width=0.8\textwidth]{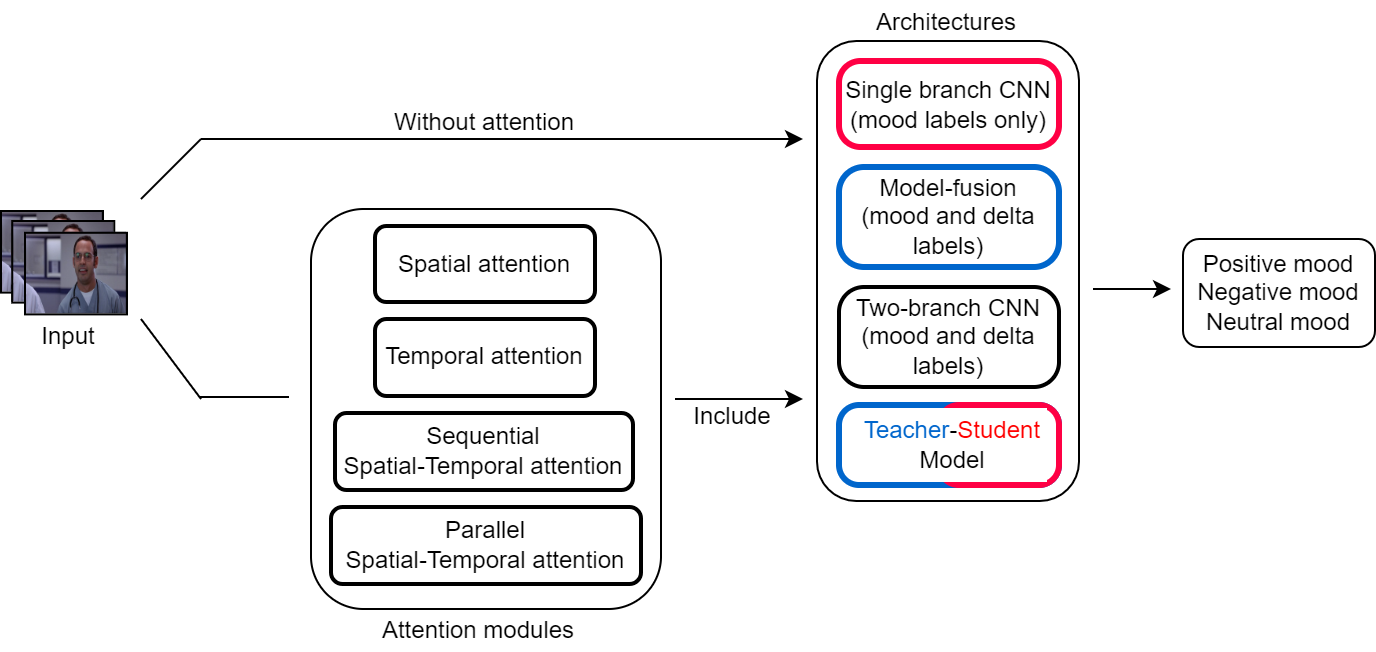} 
    \vspace{-2mm}
    \caption{Overview of the proposed mood inference framework including the various attention modules.} \vspace{-4mm}
    \label{fig:overview}
\end{figure*}

This study explores the use of emotion information for mood prediction on the AFEW-VA dataset~\cite{kossaifi2017afew} (subset of AFEW~\cite{dhall2012collecting}) comprising valence annotations. \emph{Valence} denotes the extent of pleasure or displeasure elicited by an event, and we derive \textit{positive}, \textit{negative} and \textit{neutral} mood labels from valence annotations. We additionally develop emotion change ($\Delta$) labels, which denote the change in emotion over a specific window size. For mood prediction, we preliminarily investigate (1) a single-branch 3-dimensional Convolution Neural Network (3D CNN) which uses only mood labels, (2) a two-branch (mood-$\Delta$) convolutional neural network with feature fusion followed by a Multi-Layer Perceptron, (3) a two-branch (mood-$\Delta$) convolutional neural network with end-to-end training, and (4) a teacher-student network which uses both mood and $\Delta$ labels. Empirical results described in our prior work~\cite{narayana2022improve} reveal that emotion change information is beneficial, and enhances mood prediction performance. This work additionally explores the role of attention in mood prediction. We integrate spatial attention, temporal attention, and sequential/parallel spatial-temporal (SST/PST) attention modules with the above architectures to perform mood prediction. Furthermore, to examine generalisabilty, our trained models are evaluated on the EMMA dataset~\cite{katsimerou2016crowdsourcing}, which includes mood and emotion annotations. Fig.~\ref{fig:overview} describes our framework, including the prediction architectures and attention modules. Our research contributions are summarised below: 
\begin{itemize}[noitemsep,topsep=0pt]
    \item Incorporating emotional change information is beneficial and improves mood prediction performance as compared to models trained only with mood labels.  
    \item Mood prediction improves when spatial, temporal, sequential and parallel spatial-temporal attention modules are integrated within the inference architectures.
    \item Our mood inference framework is generalizable as high prediction accuracies are achieved on EMMA~\cite{katsimerou2016crowdsourcing} by models trained on AFEW-VA~\cite{kossaifi2017afew}. Performance is evaluated at both the \textit{video} and \textit{chunk} levels, where a \textit{chunk} denotes a snippet of five contiguous frames extracted from the original video.
\end{itemize}

The paper is organised as follows. Sec.~\ref{sec:related_work} examines the literature on mood and emotion. Sec.~\ref{sec:materials} details the datasets used and the process of computing the mood and $\Delta$ labels. Mood inference models and attention modules employed for mood prediction are discussed in Sec.~\ref{sec:mood_class}. Empirical results are discussed in Sec.~\ref{sec:results}, and Conclusions presented in Sec.~\ref{sec:conclusion}. 

%
%
\section{Related Work} \label{sec:related_work}

We examine the literature in terms of (a) psychology studies relating mood and emotion, and (b) computational studies examining emotion and mood inference.  

\subsection{Mood, Emotion and their Interplay} \label{subsec:moodemo_interplay}
Psychologists consider \textit{mood} and \textit{emotion} as closely associated, but also distinguish them in terms of duration, intensity, stability, and attributes. Mood is described as a prolonged affect, lasting from minutes to hours or even days, with no specific trigger~\cite{russell2003core}. Conversely, emotions are considered to be quite brief, lasting from a few seconds to minutes, and involve the dynamic phases of onset-apex-offset~\cite{parkinson1996changing}. 
Also, while at least some emotions (known as Ekman's basic or universal emotions) are known to manifest via unique facial expressions~\cite{ekman1984expression}, moods are not known to have unique manifestations. {Cause} and {duration} primarily distinguish the two affective states~\cite{schuller2012avec}. 

Despite differences between the two affective states, the psychology literature also agrees on the confluence of the two. Ekman claims that, partially, mood is inferred from emotional signals associated with mood~\cite{ekman1999basic}. \textit{E.g.}, we may deduce that someone is in a cheerful mood due to their joyful behaviour over a period of time. Mood is also known to be instantiated by emotions with varying intensity~\cite{morris2012mood,oatley1987towards}. Furthermore, mood acts as a bias while emotions are elicited, reinforcing similar emotions, while dampening dissimilar ones~\cite{oatley1987towards}. This is similar to the mood-congruity effect, which states that positive mood facilitates the inference of mood-congruent (positive) emotions, but hampers the inference of mood-incongruent (negative) emotions, and vice-versa~\cite{schmid2010mood}. The mood-emotion loop theory proposed in~\cite{wong2016mood} states that mood is a higher level variable activating lower level latent states, such as emotions and beliefs. This study also regards mood and emotions as mechanisms, which form a loop and repeatedly trigger the arousal of each other. To summarise, though mood and emotions are regarded as distinct affective states, the psychology literature identifies a relationship between the two~\cite{morris1992functional, oatley1987towards}. 

\subsection{Mood \& Emotion Classification}\label{subsec:class_app}
Most studies focus on devising a computational framework for inferring emotions, while only a few examine mood from a computational perspective~\cite{katsimerou2015predicting}. Eye-tracking is utilized to observe that mood influences information processing styles~\cite{schmid2011mood}. Positive mood results in a global information processing style, while negative mood influences attention to detail~\cite{shapiro2013understanding}. Upper body behavior such as eye contact, arm openness, shoulder orientation, and head movement patterns are observed to perform mood inference in~\cite{thrasher2011mood}. Participants maintained a vertical head position for longer in a positive mood, as compared to negative mood. Mood prediction from recognised emotions shows that clustered emotions in the valence-arousal space\footnote{Arousal denotes the level of physiological excitation ranging from \textit{bored} to \textit{excited} following an event.} predict single mood better than multiple moods in a video~\cite{katsimerou2015predicting}.

Machine learning algorithms are popularly used for affect inference due to their ability to learn high-level representations from input features. A preliminary step in mood inference is to compile a dataset with mood annotations. Most affective databases comprise continuous and dimensional emotion annotations, \eg, HUMAINE~\cite{douglas2007humaine}, SEMAINE~\cite{mckeown2011semaine}, AFEW~\cite{kossaifi2017afew}, \etc. In these corpora, emotions are expressed via facial expressions, bosy postures, and physiological signals. EMMA~\cite{katsimerou2016crowdsourcing} is an affective database with mood labels annotated via crowdsourcing. However, EMMA is intended for monitoring the affect of care center residents, with a focus on detecting negative mood.   

Given the complexity of emotions and how they manifest, a single modality becomes inadequate for learning and representing emotions. Hence, fusing multiple modalities for emotional inference has proved to be more effective than employing unimodal methods~\cite{poria2017review, d2012consistent, liu2019multimodal}. Early fusion aims to integrate features extracted from multiple modalities \cite{koelstra2011deap}. Whereas, late fusion allows for adopting a modality-specific model, and fuses individual predictions by weighting or voting~\cite{huang2017continuous}. Hybrid fusion corresponds to a trade-off between early fusion and late fusion, and aims at exploiting their advantages in a unified framework~\cite{nemati2019hybrid}. Knowledge Distillation (KD) is the process of distilling knowledge from an ensemble of models into a single model~\cite{hinton2015distilling}. This technique achieves both model compression and learning with privileged information \cite{lopez2015unifying}. A large teacher network is trained with privileged information and transfers this knowledge to the smaller student network to achieve better inference efficiency. It is shown that knowledge can be effectively distilled using the soft target distribution produced by the teacher~\cite{hinton2015distilling}. KD is employed for facial expression recognition, with the teacher having access to fully visible faces, and the student accessing occluded faces in~\cite{georgescu2021teacher}. For micro-expression recognition, a multi-task, multi-label network is designed with a residual network as the teacher and a two-layer CNN as the student~\cite{sun2020dynamic}.

\subsection{Attention for affect classification} \label{subsec:att_for_mood}
Inspired by the way in which human perception focuses on certain parts of a given scene to better process information, the attention mechanism seeks to learn salient parts in the input data. Attention is a dynamic selection process that is realized by adaptively weighting features based on their relative importance in the input~\cite{guo2022attention}. Introduced in neural machine translation \cite{bahdanau2014neural}, attention has been widely implemented in several visual tasks like face recognition~\cite{wang2020hierarchical}, action recognition~\cite{zhu2019cuboid}, pose estimation~\cite{chu2017multi}, emotion inference~\cite{ghaleb2020multimodal}, \etc. Attention methods include spatial attention (\emph{where} to attend), channel attention (\emph{what} to attend), temporal attention (\emph{when} to attend), and hybrid approaches.  

For image classification and object detection tasks, Convolutional Block Attention Module (CBAM) is a simple-yet-efficient module implemented on feed-forward CNNs, with channel and spatial attention~\cite{woo2018cbam}. Temporal attention modules are introduced in Long-Short Term Memory (LSTM) networks to emphasize the expressive speech parts in speech emotion inference~\cite{hsiao2018effective}. Temporal attention is used to weigh audio-visual time-windows that span short video clips, as temporal attention is a powerful approach for sequence modelling, and can be used to fuse audio-visual cues over time~\cite{ghaleb2020multimodal}. To extract discriminative features from EEG signals and improve emotion inference, authors in~\cite{tao2020eeg} employ an attention-based convolutional recurrent neural network (CRNN), which adopts a channel-wise attention mechanism to adaptively assign weights to different channels. 

While attention has been employed for emotion inference tasks, less has been done towards implementing attention for mood inference. Temporal attention is integrated with a CNN model to highlight speech responses in unipolar and bipolar disorder detection~\cite{huang2019attention}. Attention-based models are found to improve mood disorder detection accuracy by roughly 11\%.

\subsection{Literature summary and research gaps}
A close examination of related work reveals that (a) whilst extensive research has focused on emotion inference, very few studies have examined mood prediction or modeled the mood-emotion interplay in their framework; (b) Labeled datasets for mood prediction are sparse, as most affective databases only include continuous/categorical emotion labels; (c) Attention mechanism has been employed largely for emotion inference, but not for mood prediction. 

Differently, this work explores (i) the utility of incorporating emotion change information for mood classification into the neutral, positive and negative classes, (ii) integrating various arrangements of spatial and temporal attention modules with mood prediction architectures, (iii) model generalizability by evaluating AFEW-VA \cite{kossaifi2017afew}-trained models on the EMMA \cite{katsimerou2016crowdsourcing} dataset, and (iv) evaluating mood prediction performance at the chunk and video levels. Our results demonstrate that accounting for the mood-emotion interplay,  by training a teacher-student network with both mood and $\Delta$ labels, enables better mood inference than mood-based models. Incorporating attention in this framework further improves prediction performance and model generalizability. 

\section{Materials} \label{sec:materials}
This section presents datasets used in our study, followed by the procedure adopted for synthesizing mood labels from valence annotations.

\begin{figure*}[!ht]
    \centering
    \includegraphics[width=0.6\textwidth]{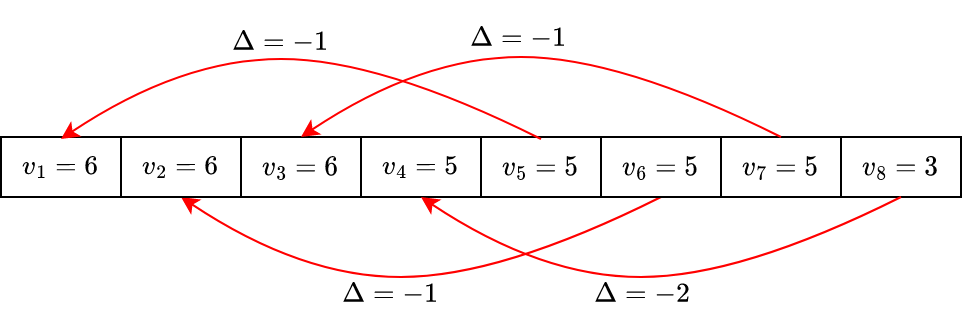} 
    \vspace{-2mm}
    \caption{Illustration of the $\Delta$ label computation process. $v_i$ denotes the valence of the $i^{th}$ frame.} \vspace{-5mm}
    \label{fig:delta}
\end{figure*}

\subsection{Datasets} \label{subsec:datasets}
We employ the AFEW-VA dataset~\cite{kossaifi2017afew} for compiling mood labels. We further evaluate the trained models on the EMMA dataset~\cite{katsimerou2016crowdsourcing}. The two datasets are described below:
\subsubsection{\textbf{AFEW-VA}} AFEW~\cite{dhall2012collecting} is a corpus containing facial expression videos. AFEW-VA~\cite{kossaifi2017afew} is a subset of AFEW containing time-continuous valence-arousal annotations. AFEW-VA clips are 10--145 frames long, and are captured under challenging indoor and outdoor conditions. Each video frame is annotated for valence and arousal intensities in the [-10,10] range. The annotations are produced by two expert annotators (one male, one female). 

\subsubsection{\textbf{EMMA}} EMMA~\cite{katsimerou2016crowdsourcing} is an affective database containing 180 acted non-interactive videos typically depicting daily scenarios, and subtle facial and body expressions. The videos are lab-recorded, and depict 15 (6 male, 9 female) actors. The situational context and scenarios are defined to the actors, followed by mood induction procedures to elicit the target mood. Each video is associated with a categorical mood label acquired via crowdsourcing. 

\subsection{Labels}\label{subsec:labels}
\subsubsection{\textbf{Mood labels}}
For each AFEW-VA video, we assigned a \textit{mood label} based on the per-frame valence level. We assigned a model label of +1 (\emph{positive} mood) to frames with valence values in the range (3,10], 0 (\emph{neutral} mood) to the valence range [-3,3], and -1 (\emph{negative} mood) to the (-3,-10] range. We then synthesized video clips/input samples from the original video considering overlapping sequences of 5 contiguous frames. Each clip was assigned the mode of the per-frame mood labels in a sequence. Via this scheme, we extracted a total of 27,651 clips from the AFEW-VA dataset. 

As EMMA already contains mood labels, we again consider overlapping five-frame sequences from each video, resulting in a total of 536,063 clips. The mood label for each EMMA sample is the same as the source video. 

\subsubsection{\textbf{Emotion change ($\Delta$) labels}}

In addition to mood labels, each input sample is assigned an \emph{emotion change} or \emph{$\Delta$ label}, which refers to the change in emotional valence over $k$ frames. For a video with $n$ frames and a window size of $k$, it is computed as the difference in the valence between the $t^{th}$ frame and $(t - k + 1)^{th}$ frame, for $t = k, k+1, ..., n$. For example, considering $k = 5$, if $v_5 = -4$ and $v_1 = -2$, where $v_5$ and $v_1$  respectively denote the valence of the fifth and the first frames, $\Delta = v_5 - v_1 = -2$ (See Fig.~\ref{fig:delta}). We assign the sample $\Delta$ label as $\sgn(\Delta)$ defined below: 
\begin{equation}
\sgn \Delta =\begin{cases} 
-1 & \text{if } \Delta < 0, \\
0 & \text{if } \Delta = 0, \\
1 & \text{if } \Delta > 0. \end{cases}
\end{equation}

Each AFEW-VA sample has mood and $\Delta$ labels $\in \{0, -1, +1\}$, which are employed for the model training. Since EMMA is used only for testing, mood labels alone suffice for the EMMA samples.

\section{Mood classification Architectures} \label{sec:mood_class}

\begin{figure*}[!htbp]
    \centering
    \includegraphics[width=0.8\textwidth]{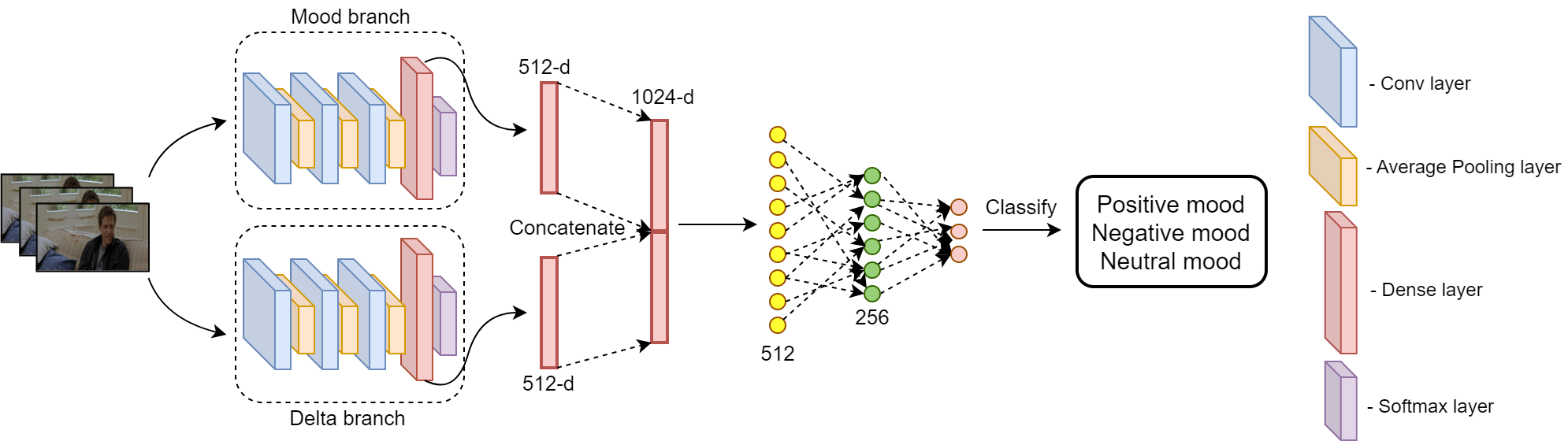} \vspace{-2mm}
    \caption{Architecture of 2-CNN+MLP fusing mood and emotion change ($\Delta$) information. The dashed rectangles in each branch show the 1-CNN, and the top and bottom branches are trained via the mood and $\Delta$ labels. Legend on the right side denotes color codes for the various CNN layers.} 
    \label{fig:2-CNN+MLP}
    \centering
    \includegraphics[width=0.8\textwidth]{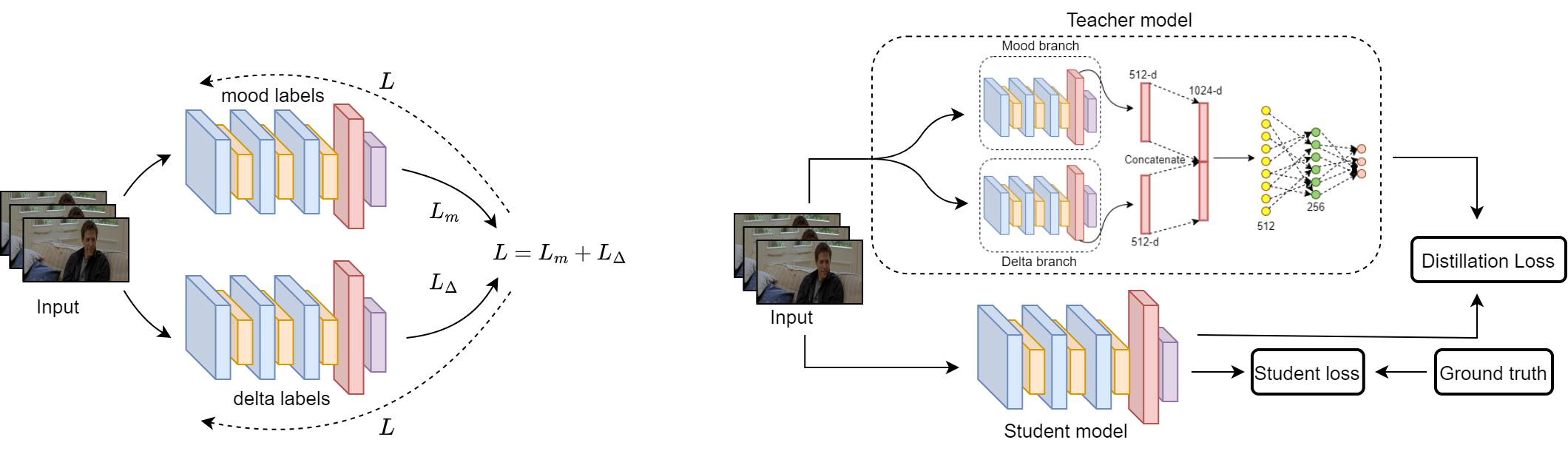} 
    \vspace{-2mm}
    \caption{\textbf{(Left)} Architecture of 2-CNN which is composed of two 1-CNN models. \textbf{(Right)} Architecture of TS-Net whose layers are as described in Fig~\ref{fig:2-CNN+MLP}.} 
    \label{fig:2-CNN+TS}
    \centering
    \includegraphics[width=0.85\textwidth]{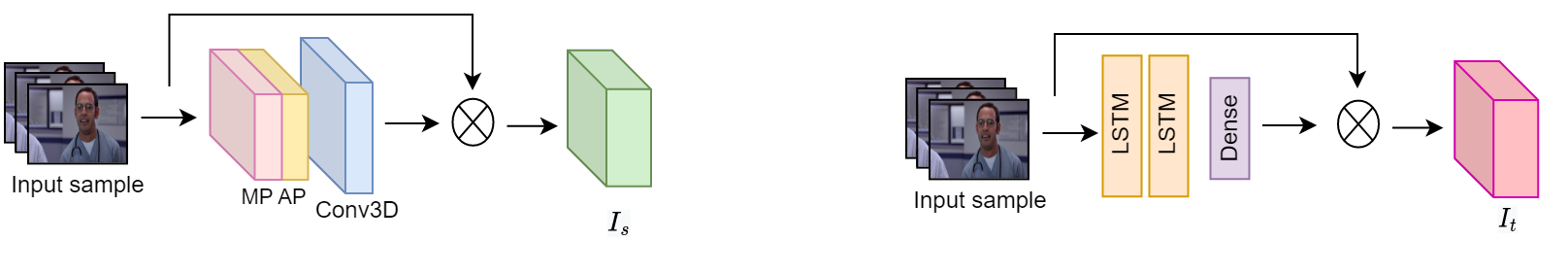} 
    \vspace{-6mm}
    \caption{\textbf{(Left)} Architecture of the spatial attention module. \textbf{(Right)} Architecture of the temporal attention module.}\vspace{-4mm}
    \label{fig:sp_temp_att}
\end{figure*}

As we seek to explore if the use of emotion information is beneficial for mood classification, we train different deep neural networks with mood and $\Delta$ labels, and compare their performance against models trained only with mood labels. The trained model architectures are described below followed by a discussion of the attention modules. 

\subsection{Classification Architectures}\label{subsec:base}
\subsubsection{\textbf{1-CNN}} We employ a single branch three-layered convolutional neural network, or 1-CNN for mood classification. The model has three convolutional layers which comprise of 16, 32, and 32 kernels respectively with size $3 \times 3 \times 3$ and each of them convolve the input sample with a stride of 3. Followed by each of the convolutional layer is an average pooling layer with a stride of 2-pixel regions. The output of the last convolutional layer is flattened, followed by batch normalization. Following the dense layer comprising 512 neurons is the Softmax layer with three neurons corresponding to the mood classes. 

The input dimensionality for the 1-CNN is $5 \times 32 \times 32 \times 3$, with each sample comprising 5 frames of size $32 \times 32$ and 3 channels. The model is optimized with the categorical cross-entropy loss function, with the Adam optimiser used for stochastic gradient descent. Fine-tuned hyper-parameters include learning rate $\in$ $\{10^{-3}, 10^{-5}\}$, batch size $\in$ $\{64, 128, 256\}$, and dropout rate $\in$ \{0.4, 0.5\}.

\subsubsection{\textbf{2-CNN + MLP}}
Multimodal systems have gained increasing attention from researchers~\cite{cimtay2020cross, praveen2022joint, alghowinem2016multimodal} by overcoming limitations of unimodal counterparts~\cite{d2012consistent, d2015review}. Multimodal systems fuse complementary information from different modalities at multiple levels. To examine the influence of emotion change ($\Delta$) on mood prediction, we fuse the two employing 2-CNN + Multi-layer Perceptron (MLP) as in Fig.~\ref{fig:2-CNN+MLP}. The 2-CNN + MLP network is a two-branch model, with two identical 1-CNNs along each branch; mood labels are fed to one branch, and $\Delta$ labels fed to the other. From the penultimate layer in each branch, a 512-dimensional vector is obtained, and the two vectors are concatenated to obtain a 1024-feature vector input to the MLP for mood classification. $\Delta$ labels are assumed to be available during the train and test phases in this model. Dimensionality of the input samples and hyper-parameters are identical to 1-CNN. 

\subsubsection{\textbf{2-CNN}}
As a substitute to model fusion, we trained an end-to-end model for mood classification~\cite{liu2021end,trigeorgis2016adieu} via a 2-CNN model shown in Fig.~\ref{fig:2-CNN+TS}. Similar to the 2-CNN + MLP, the top and bottom branches are trained with mood and $\Delta$ labels respectively, but both branches are trained simultaneously but not individually. The categorical cross-entropy losses $L_m$ and $L_{\Delta}$ from the two branches are summed up, and the cumulative loss is minimised. Unlike the 2-CNN + MLP, where $\Delta$ labels are used both in train and test phases, $\Delta$ labels act as auxiliary information for the 2-CNN model, and are required only during the training phase (mood labels suffice for testing).

\subsubsection{\textbf{TS-Net}}
Besides model fusion and end-to-end training, we employ Knowledge Distillation, which is the process of transferring knowledge from a large model to smaller ones \cite{hinton2015distilling}. We employ a Teacher-Student Network (Fig.~\ref{fig:2-CNN+TS}(right)) where the teacher (large model) distills knowledge to the student (smaller model), and thereby the student tries to mimic the teacher. The pre-trained teacher model (2-CNN + MLP) uses both mood labels and $\Delta$ labels in the training phase to distill knowledge to the student (1-CNN). The student model, whic only utilizes mood labels, is activated for inference.  

The Softmax layer of the student model involves the \emph{temperature (T)} hyper-parameter, which controls the smoothness of the output probabilities. As $T$ grows, the output Softmax probabilities become softer, providing more information regarding the inter-class relations than one-hot labels \cite{hinton2015distilling}. The distillation loss is computed via the Kullback–Leibler (KL) divergence measure, while the student loss is computed via sparse categorical cross-entropy. The overall loss of the TS-Net, $L_{TS}$ is computed as the weighted sum of the distillation loss $L_{dist}$ and the student loss $L_{stu}$, and is given by:
\begin{equation}
L_{TS} = \alpha L_{\text{stu}}  + (1 - \alpha) L_{\text{dis}}
\end{equation}
where $\alpha$ is a training hyper-parameter. The other fine-tuned hyper-parameters include batch size $\in \{16, 64, 128\}$, $T$ $\in \{3, 5, 7\}$ and $\alpha \in \{0.05, 0.1, 0.15, 0.2, 0.25, 0.3\}$.

\subsection{Attention modules} \label{subsec:att}

\subsubsection{\textbf{Spatial attention}}
In humans, spatial attention allows to selectively process visual information through prioritization of locations within a field. Spatial attention modules have been widely implemented in tasks like action recognition, affect classification, \etc. to improve model performance~\cite{li2018unified, jia2020sst}. Inspired by~\cite{woo2018cbam}, we generate spatial attention maps to emphasize meaningful features. As shown in Fig.~\ref{fig:sp_temp_att} (left), spatial attention is computed by applying max and average-pooling operations along the channel axis, followed by concatenation to generate a feature descriptor. A 3D convolutional layer is applied on the concatenated feature, and the output is multiplied with the input to generate a spatial attention map, which emphasizes \emph{where} to look for in the video clip. 

\subsubsection{\textbf{Temporal attention}}
Temporal attention is a dynamic mechanism which determines \emph{when} to pay attention, and directs visual focus to specific time instants. A video represents a sequence of visual segments with large content variance and complexity, where the amount of valuable information provided by frame segments is generally unequal~\cite{sigurdsson2017asynchronous}, and discriminative information is provided by only some key frames. To model temporal dynamics in video, we apply LSTM networks, thereby capturing long-term temporal information. As shown in Fig \ref{fig:sp_temp_att} (right), we use two LSTM layers with 128 hidden units per layer, followed by a dense layer with a single neuron. The input feature is multiplied with the output of the dense layer, facilitating the enhanced model to distinguish the most informative frame(s) for mood classification. 

\subsubsection{\textbf{Integrating attention modules}}
\begin{figure*}[!ht]
    \centering
    \includegraphics[width=0.9\textwidth]{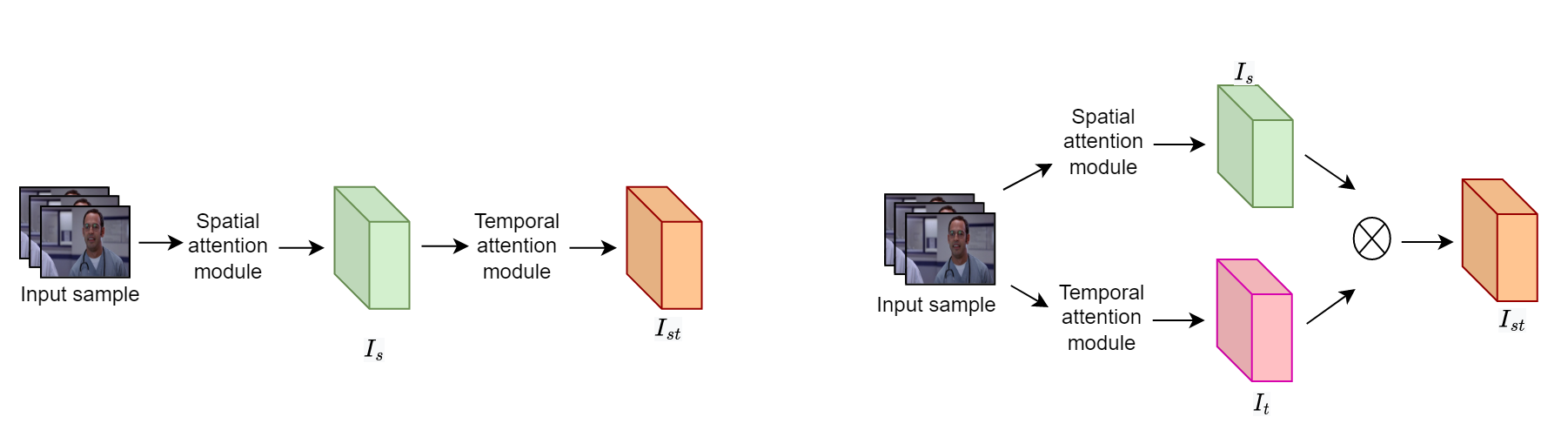} 
    \vspace{-4mm}
    \caption{\textbf{(Left)} Sequential Spatial-Temporal (SST) attention module. \textbf{(Right)} Parallel Spatial-Temporal (PST) attention module.}\vspace{-2mm}
    \label{fig:sp_temp_ser_para}
\end{figure*}

Given an input clip, the spatial and temporal attention modules generate complementary attention information by respectively identifying salient regions and key frames. We integrate spatial and temporal attention modules into the classification networks, and also arrang these modules sequentially and parallely to synthesize the \emph{sequential} spatial-temporal (SST) and \emph{parallel} spatial-temporal (PST) attention architectures. 
\begin{itemize}
    \item \textbf{SST Attention:} In SST Attention (Fig.~\ref{fig:sp_temp_ser_para} (left)), the input sample $I$ is sequentially fed into the spatial followed by the temporal attention modules. $I$ is input to the spatial module first, and the attention map $F_s$ is then convolved with $I$ to produce $I_s$ which is fed to the temporal module. The temporal map $F_t$ is then multiplied with $I_s$ to obtain $I_{st}$, which is fed to the classifier networks.
    \item \textbf{PST Attention:} In PST Attention (Fig.~\ref{fig:sp_temp_ser_para} (right), the input sample $I$ is fed parallelly to the spatial and temporal attention modules. The spatial ($F_s$) and temporal ($F_t$) attention maps are then convolved with $I$ to respectively obtain $I_s$ and $I_t$, which are multiplied to obtain $I_{st}$ input to the classification models. 
\end{itemize}

\subsection{Integration with classification models} \label{subsec:integration}

We integrate the spatial and temporal attention modules individually and in combination with each of the mood classifer networks, as shown in Fig.~\ref{fig:overview}. Corresponding results are presented in Sec~\ref{sec:results}.

\subsection{Evaluating on EMMA dataset} \label{subsec:test_emma}
As mentioned in Sec \ref{subsec:datasets}, EMMA is an affective database comprising acted videos with mood and emotion annotations. There are $\approx 200$ frames in each EMMA video, while the maximum video frames in AFEW-VA is 145. However, EMMA is compiled under lab conditions and with actors, unlike AFEW-VA which represents an \emph{in-the-wild} video dataset. To examine if (a) the mood and $\Delta$ labels assigned over smaller (5-frame) AFEW-VA snippets were applicable over larger EMMA clips, and (b) the mood prediction models trained on AFEW-VA were generalizable to EMMA, we evaluated all mood classifiers on EMMA utilizing only the mood labels. 




On performing a 90-10 stratified and random train-test split of the EMMA dataset, the test set comprised 18 videos, with 11 negative, 3 positive, and 4 neutral mood labels. Identical to AFEW-VA, each EMMA clip is of dimensionality $5 \times 32 \times 32 \times 3$, resulting in a total of 62,410 samples. We evaluate all trained models, with and without integrating attention modules, on EMMA. 
While evaluating models on EMMA, each 5-frame sample or \emph{chunk} from the original video is tested, and we then evaluate models at both the \emph{chunk} and \emph{video} levels. This enables comparison of mood prediction performance over longer (video-level) and shorter (chunk-level) episodes depicting visual behavior. 

\subsection{Performance metric} 
Subject-independent 5-fold cross validation is used to evaluate all models to eliminate leakage effects. \textit{Mean accuracy} over the five folds is reported as our performance metric. Results evaluating performance on the EMMA dataset also report the average performance over the five training models synthesized over the cross-validation process.

\section{Results and discussion} \label{sec:results}

\begin{table*}[!t]
\caption{Performance comparison without and with integrating ttention modules on the \textbf{AFEW-VA} dataset.}\vspace{-2mm}
\label{tab:with_without_att}
\centering
\renewcommand{\arraystretch}{1.2}
\resizebox{0.8\textwidth}{!}{%
\begin{tabular}{@{}lccccc@{}}
\toprule
\multirow{4}{*}{\textbf{Classification model}} & \multirow{4}{*}{\textbf{No attention}} & \multicolumn{4}{c}{\textbf{With attention}} \\ \cmidrule(l){3-6} 
 &  & \multicolumn{1}{c}{\textbf{Spatial}} & \multicolumn{1}{c}{\textbf{Temporal}} & \multicolumn{1}{c}{\begin{tabular}[c]{@{}c@{}}\textbf{Spatial-Temporal}\\ \textbf{(Sequential)}\end{tabular}} & \begin{tabular}[c]{@{}c@{}}\textbf{Spatial-Temporal}\\ \textbf{(Parallel)}\end{tabular} \\ \midrule
\begin{tabular}[c]{@{}c@{}}1-CNN (Mood)\end{tabular} & 0.35 $\pm$ 0.10 & \multicolumn{1}{c}{0.68 $\pm$ 0.18} & \multicolumn{1}{c}{0.73 $\pm$ 0.16} & \multicolumn{1}{c}{0.65 $\pm$ 0.11} & 0.70 $\pm$ 0.08 \\
\begin{tabular}[c]{@{}c@{}}1-CNN ($\Delta$)\end{tabular} & 0.53 $\pm$ 0.11 & \multicolumn{1}{c}{0.47 $\pm$ 0.10} & \multicolumn{1}{c}{0.40 $\pm$ 0.13} & \multicolumn{1}{c}{0.62 $\pm$ 0.10} & 0.56 $\pm$ 0.08 \\
2-CNN + MLP & 0.87 $\pm$ 0.15 & \multicolumn{1}{c}{0.66 $\pm$ 0.04} & \multicolumn{1}{c}{0.46 $\pm$ 0.08} & \multicolumn{1}{c}{0.78 $\pm$ 0.14} & 0.63 $\pm$ 0.06\\
2-CNN & 0.70 $\pm$ 0.07 & \multicolumn{1}{c}{\textbf{0.91 $\pm$ 0.02}} & \multicolumn{1}{c}{\textbf{0.91 $\pm$ 0.02}} & \multicolumn{1}{c}{\textbf{0.91 $\pm$ 0.01}} & \textbf{0.90 $\pm$ 0.02} \\
TS-Net & \textbf{0.89 $\pm$ 0.10} & \multicolumn{1}{c}{\textbf{0.85 $\pm$ 0.09}} & \multicolumn{1}{c}{\textbf{0.80 $\pm$ 0.08}} & \multicolumn{1}{c}{\textbf{0.83 $\pm$ 0.09}} & \textbf{0.86 $\pm$ 0.06} \\
\bottomrule
\end{tabular}%
}
\end{table*}

\subsection{Classification Experiments}

\subsubsection{Mood inference without attention modules}
Table~\ref{tab:with_without_att} presents AFEW-VA mood classification results without and with various attention modules. The first column, which tabulates accuracies excluding attention modules, conveys that the 1-CNN (Mood) trained with mood labels performs worst, revealing that mood prediction is a non-trivial problem. As a pre-cursor to the 2-CNN + MLP and 2-CNN models which include a CNN-branch trained with $\Delta$ labels, we also evaluated a 1-CNN ($\Delta$) model trained and tested with $\Delta$ labels; this model performs better than 1-CNN (Mood) implying that emotion changes are easier detected than mood.  

Comparing the performance of the 2-CNN + MLP and 2-CNN networks, both of which are trained with both mood and $\Delta$ labels, we note that the 2-CNN + MLP network performs considerably than the 2-CNN whilst both achieve superior mood inference vis-\'a-vis the 1-CNN (Mood) model. These findings cumulatively reveal that including emotion-change information for mood prediction, either implicitly as auxiliary information (2-CNN) or explicitly as additional features (2-CNN + MLP), benefits mood inference. The highest mood inference accuracy of 0.89 is obtained with the TS-Net, where the 2-CNN + MLP denotes the pre-trained \textit{teacher} model trained with both mood and auxiliary $\Delta$ labels, while 1-CNN denotes the \textit{student} model, trained only with mood labels. 

Overall, these results affirm our claim that modeling the mood-emotion interplay and incorporating emotion change information for mood inference enables superior mood prediction as compared to utilizing mood information alone. $\Delta$ labels capture emotion changes over a short duration (local phenomenon), in comparison to mood labels which are applicable over a longer duration (global phenomenon). Our results reveal that short-term emotion changes play a crucial role in inferring long-term mood.

\subsubsection{Mood inference with attention}
Mood prediction results on integrating attention modules to the classification models are shown in Table~\ref{tab:with_without_att} (`With attention' columns). We examine if integrating attention  into the classification architecture in addition to learning emotion changes improves mood prediction. A horizontal comparison in Table \ref{tab:with_without_att} conveys that irrespective of the attention module, the 1-CNN performance significantly improves as compared to the baseline 1-CNN trained with mood labels.
Conversely, the 2-CNN + MLP and TS-Net (with the 2-CNN + MLP as the Teacher model) do not benefit when attention modules are integrated, and their prediction accuracies somewhat reduce. The 1-CNN ($\Delta$) does not benefit with either spatial or temporal attention, but its performance improves with sequence and parallel configurations of the spatial and temporal modules.  

\begin{figure*}[!ht]
    \centering
    \includegraphics[width=0.6\textwidth, height=4.2cm]{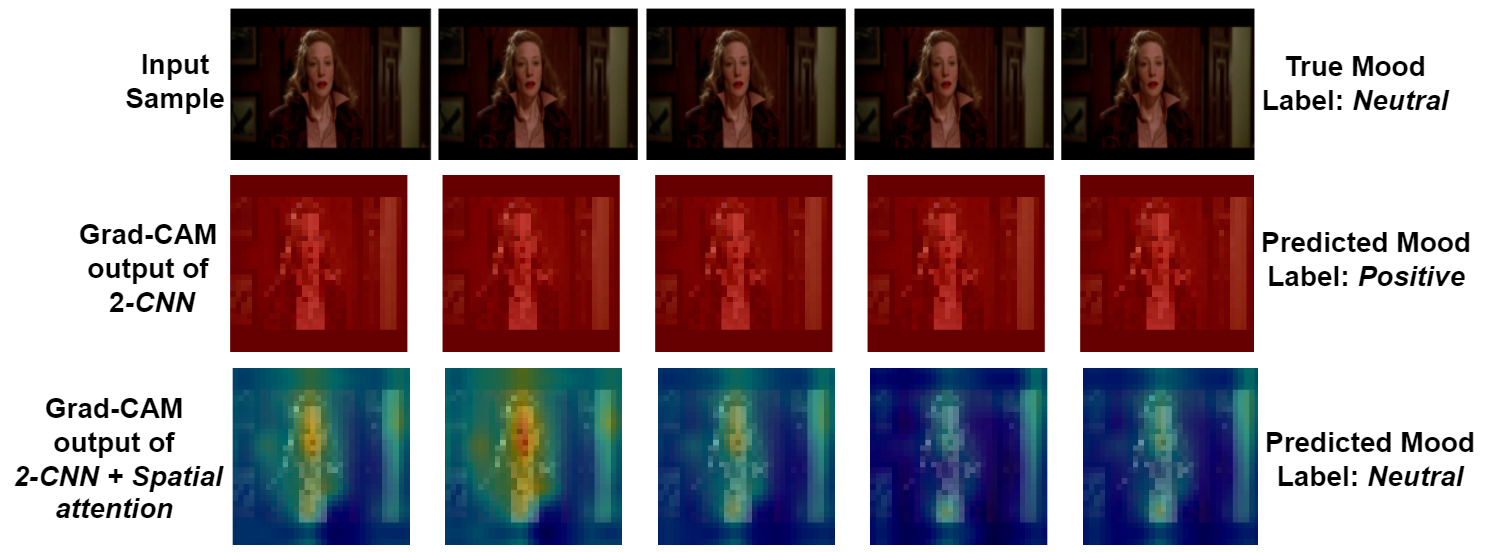} 
    \vspace{-3mm}
    \caption{\textbf{Mood prediction with spatial attention:} The first row shows an input clip with \textit{neutral} label. The second row shows GradCAM~\cite{selvaraju2017grad}  maps of the 2-CNN without attention (incorrect prediction), while the third row presents the 2-CNN + Spatial attention maps (correct prediction). (best viewed in color).}\vspace{-4mm}
    \label{fig:sp_temp_att_GRAD}
\end{figure*}

With the 2-CNN network, where $\Delta$ labels are employed as auxiliary information, a significant performance improvement is noted with each of the attention modules. This observation reveals that the attention mechanism enables the 2-CNN to focus on informative pixels and important temporal segments for mood prediction. Fig.~\ref{fig:sp_temp_att_GRAD} qualitatively supports this finding, where the input sample with a \textit{neutral} mood label is incorrectly predicted as \textit{positive} by the 2-CNN without any attention modules. Integration of the spatial attention module enables the classification model to predominantly focus on facial regions and achieve the correct prediction. Overall, empirical results reveal that classification models inherently learning from emotion change information (1-CNN ($\Delta$), 2-CNN +MLP and TS-Net) do not additionally benefit from attention, while mood-based networks immensely benefit as attention facilitates focus on the facial region.   

To investigate the influence of emotion change ($\Delta$) information on mood prediction, we compare Table~\ref{tab:with_without_att} results vertically. With spatial attention, while the 2-CNN and TS-Net perform better than the 1-CNN, the results are not significantly different. Likewise, integrating temporal attention also results in a higher accuracy with the 2-CNN and TS-Net as compared to 1-CNN (Mood). With respect to serial and parallel combinations of the spatial and temporal modules, with SST attention, the 2-CNN achieves significantly higher performance than 1-CNN (Mood) as per a two-sample $t$-test ($t(8)=-4.7, p < 0.05$). A significant difference is also observed between TS-Net and 1-CNN networks ($t(8)=-2.6, p < 0.05$). PST attention also yields significantly higher results with the 2-CNN ($t(8)=-5.2, p<0.05$), and the TS-Net ($t(8)=-3.4, p<0.05$) as compared with the 1-CNN. 

\begin{figure*}[!ht]
    \centering
    \includegraphics[width=0.6\textwidth, height=4.2cm]{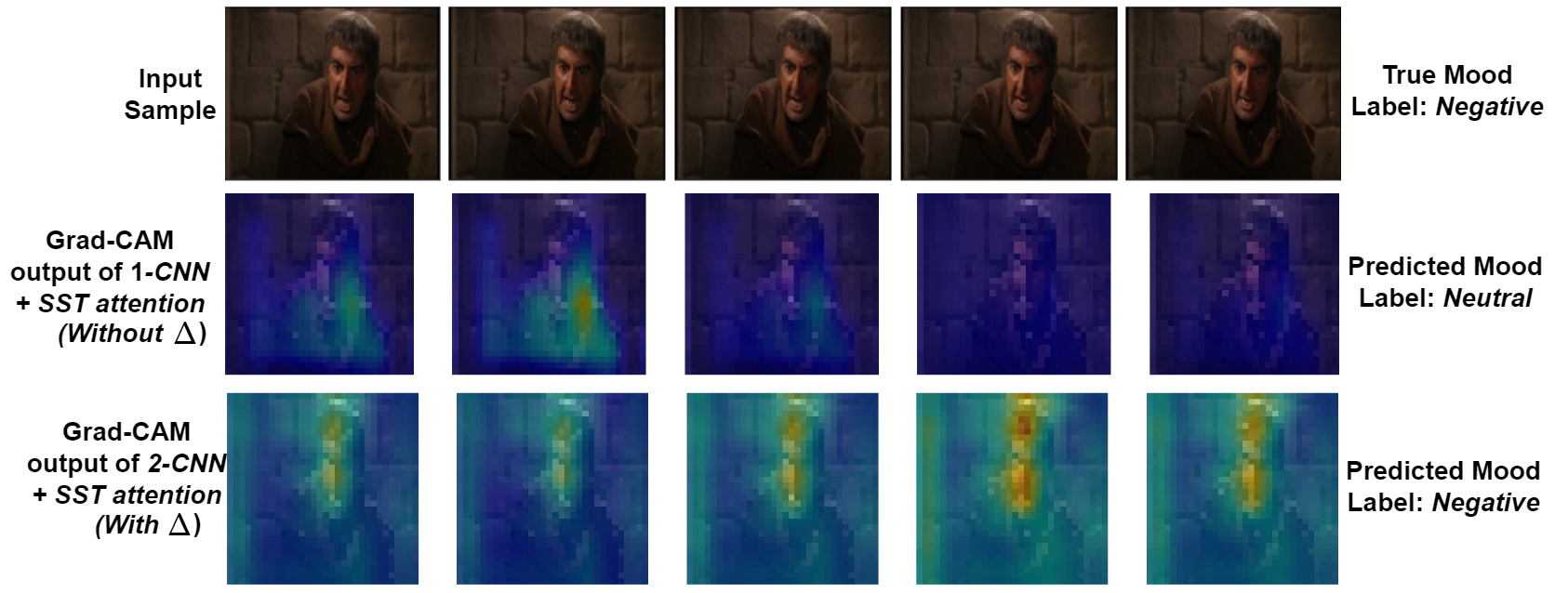} 
    \vspace{-3mm}
    \caption{\textbf{Mood prediction with $\Delta$ labels and SST attention:} GradCAM maps depicting activation maps for a \textit{negative} mood clip (first row). The second row shows GradCAM maps for 1-CNN + SST (incorrect mood prediction), while the third row shows activation maps for the 2-CNN + SST network with the correct mood prediction. Incorporating $\Delta$ information for mood learning is beneficial, and enables the  2-CNN to focus on the face better than the 1-CNN. (best viewed in color).}
    \label{fig:sp_temp_ser_seq_GRAD}\vspace{-4mm}
\end{figure*}

As mentioned above, the attention modules and learning from emotion change information are complementary means to improve mood recognition performance. A qualitative depiction of how learning information pertaining to emotion change is shown in Fig.~\ref{fig:sp_temp_ser_seq_GRAD}. Attention maps for the input sample shown in the first row obtained with the 1-CNN (Mood) network including SST attention (2nd row) show that the model focuses regions in the vicinity of the face for mood inference. However, adding auxiliary information in the form of emotion change labels (3rd row) strongly localizes focus on the facial region to achieve superior mood inference. Cumulatively, Fig.~\ref{fig:sp_temp_att_GRAD} and Fig.~\ref{fig:sp_temp_ser_seq_GRAD} reinforce the observation that incorporating attention or emotion-change information for mood inference, are complementary means to the same end.

\subsubsection{Evaluation on EMMA}
\begin{table*}[!t]
\caption{Performance comparison on \textbf{AFEW-VA} and \textbf{EMMA}. EMMA test results include without and with attention modules.}\vspace{-2mm}
\label{tab:afew-va_emma}
\centering
\renewcommand{\arraystretch}{1.2}
\resizebox{0.8\textwidth}{!}{%
\begin{tabular}{@{}lcccccc@{}}
\toprule
\multirow{4}{*}{\textbf{Classification model}} & \multirow{4}{*}{\textbf{AFEW-VA}} & \multirow{4}{*}{\begin{tabular}[c]{@{}c@{}}\textbf{EMMA}\\ \textbf{(No attention)}\end{tabular}} & \multicolumn{4}{c}{\begin{tabular}[c]{@{}c@{}}\textbf{EMMA}\\ \textbf{(With attention)}\end{tabular}} \\ \cmidrule(l){4-7} 
 &  &  & \multicolumn{1}{c}{\textbf{Spatial}} & \multicolumn{1}{c}{\textbf{Temporal}} & \multicolumn{1}{c}{\begin{tabular}[c]{@{}c@{}}\textbf{Spatial-Temporal}\\ \textbf{(Sequential)}\end{tabular}} & \begin{tabular}[c]{@{}c@{}}\textbf{Spatial-Temporal}\\ \textbf{(Parallel)}\end{tabular} \\ \midrule
1-CNN & 0.35 $\pm$ 0.10 & 0.41 $\pm$ 0.17 & \multicolumn{1}{c}{0.21 $\pm$ 0.01} & \multicolumn{1}{c}{0.20 $\pm$ 0.10} & \multicolumn{1}{c}{0.21 $\pm$ 0.11} & 0.24 $\pm$ 0.11 \\
2-CNN & 0.70 $\pm$ 0.07 & 0.19 $\pm$ 0.00 & \multicolumn{1}{c}{0.21 $\pm$ 0.03} & \multicolumn{1}{c}{0.19 $\pm$ 0.00} & \multicolumn{1}{c}{0.13 $\pm$ 0.03} & 0.57 $\pm$ 0.13 \\
TS-Net & 0.89 $\pm$ 0.10 & \textbf{0.69 $\pm$ 0.00} & \multicolumn{1}{c}{\textbf{0.62 $\pm$ 0.00}} & \multicolumn{1}{c}{\textbf{0.67 $\pm$ 0.00}} & \multicolumn{1}{c}{\textbf{0.67 $\pm$ 0.01}} & \textbf{0.67 $\pm$ 0.00}\\
\bottomrule
\end{tabular}%
}
\end{table*}

\begin{table*}[!t]
\caption{EMMA results at the \textbf{chunk and video levels}, using TS-Net with and without attention modules.}\vspace{-2mm}
\label{tab:emma_video_chunk}
\centering
\renewcommand{\arraystretch}{1.2}
\resizebox{1.1\columnwidth}{!}{%
\begin{tabular}{@{}lcc@{}}
\toprule
\multirow{2}{*}{\textbf{Classification model}} & \multirow{2}{*}{\textbf{Chunk-level}} & \multirow{2}{*}{\textbf{Video-level}} \\
 &  &  \\ \midrule
TS-Net + No attention & 0.69 $\pm$ 0.00 & 0.54 $\pm$ 0.16 \\
TS-Net + Spatial attention & 0.62 $\pm$ 0.00 & {0.85 $\pm$ 0.23} \\
TS-Net + Temporal attention & 0.67 $\pm$ 0.00 & \textbf{0.89 $\pm$ 0.24} \\
TS-Net + Spatial-Temporal attention (Sequential) & 0.67 $\pm$ 0.01 & \textbf{0.89 $\pm$ 0.24} \\
TS-Net + Spatial-Temporal attention (Parallel) & 0.67 $\pm$ 0.00 & {0.83 $\pm$ 0.22}\\
\bottomrule
\end{tabular}%
}
\end{table*}

As unlike the AFEW-VA dataset where mood labels are derived from valence annotations, the EMMA dataset comprises mood annotations. Also, mood labels in EMMA correspond to longer video segments as compared to EMMA. To examine if the models trained on AFEW-VA, and specifically to investigate if incorporating emotion change information can positively impact EMMA mood prediction, we evaluated all trained models on the EMMA dataset and results are presented in Table~\ref{tab:afew-va_emma}. With no attention, the TS-Net performs best with 69\% accuracy followed by the 1-CNN, while the 2-CNN performs worst on EMMA. Including attention modules improves 2-CNN mood prediction performance, while not benefitting inference with the 1-CNN or TS-Net. Overall, the TS-Net performs best on EMMA; we did not fine-tune the AFEW-VA models on EMMA as our main objective was to test model generalizability. 



We also evaluated mood prediction performance at the chunk (5-frame episodes) and video levels on EMMA. Table~\ref{tab:emma_video_chunk} presents the corresponding results using TS-Net, which achieves optimal performance on AFEW-VA. While TS-Net without attention yields significantly higher results at the chunk level than at the video level, integrating attention modules results in significantly higher accuracies at the video-level. A highest accuracy of 0.89 is achieved with the models involving temporal attention. These results convey that while the AFEW-VA models generalize better to EMMA at the video-level, affirming the validity of our annotation and model training processes.


\subsection{Discussion Summary}
The classification experiments convey that (a) incorporating emotion change information in addition to the mood labels positively impacts mood prediction; (b) integration of various attention modules to the architectures enables models to focus on key areas/input frames and improves mood prediction; (c) Qualitative results furthermore demonstrate that attention and emotion-change information play a complementary role towards enhancing mood inference. Additionally, AFEW-VA trained models generalize better to EMMA at the video-level than at the chunk-level, confirming that our methodology for annotating mood labels is valid, and can generalize to videos of longer duration.

\section{Conclusion} \label{sec:conclusion}

With majority of the contemporary studies focusing on emotion inference using multiple modalities and attention mechanism, negligible work has been done towards mood inference and examining the interplay between mood and emotions. The applicability of attention methods for examining mood is also neglected. As a step towards fulfilling these disparities, this work proposes using emotion change information along with attention mechanism for performing mood prediction. Using the AFEW-VA dataset, a single-branch CNN, two-branch CNN with MLP, two-branch CNN with end-to-end training and teacher-student network are employed as base models for mood prediction. Emotion change ($\Delta$) labels, referring to the difference in emotion over a fixed window size, is used in addition to the mood labels for examining mood. Further, various attention mechanisms are integrated to the base models to enhance the model's performance. These models trained and tested on AFEW-VA are evaluated on EMMA dataset to explore the generalisability of approach. 

Overall, we observe that without integrating attention modules in two-branch CNN and teacher-student network, incorporating emotion change information along with the mood labels improves mood prediction performance. The integration of parallel spatial-temporal attention modules in these models also yields similar results, further validating our claim. This approach is also generalizable as identical results with teacher-student network are observed when the model is evaluated on EMMA dataset. Better results in the video-level analysis as compared to chunk-level analysis shows that the model is also robust towards videos of longer duration. Future work will investigate the use of semi-supervised and weakly supervised techniques, and substitute emotion-change labels with difference or class-activation maps.


\section{Acknowledgements}
This research is partially funded by the Australian Government through the Australian Research Council’s Discovery Projects funding scheme (project DP190101294).
\ifCLASSOPTIONcaptionsoff
  \newpage
\fi

\bibliography{references.bib}
\bibliographystyle{IEEEtran}
\vskip -2\baselineskip plus -1fil
\begin{IEEEbiography}[{\includegraphics[width=1in,height=1.25in,clip,keepaspectratio]{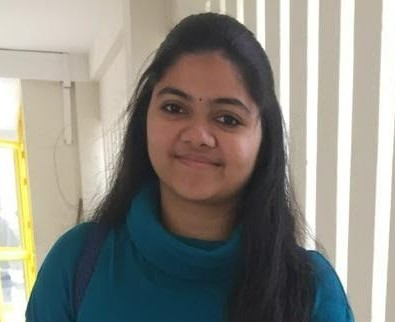}}]{Soujanya Narayana} received her Bachelor's degree from Regional Institute of Education, Mysore, India in 2015 and Masters degree from Christ University, Bengaluru, India in 2017. She is currently a PhD student in the Human Centered Technology Research Centre at the University of Canberra, Australia. Her research interests include affective computing in human-computer interaction and computer vision. She is a student member of the IEEE. \end{IEEEbiography}
\vskip -3\baselineskip plus -1fil
\begin{IEEEbiography}[{\includegraphics[width=1in,height=1.25in,keepaspectratio]{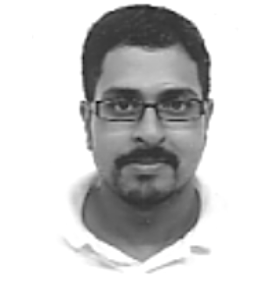}}]{Ramanathan Subramanian} received his Ph.D. in Electrical and Computer Engg. from NUS in 2008. He is Associate Professor in the School of IT \& Systems, University of Canberra. His past affiliations include IIT Ropar, IHPC (Singapore), U Glasgow (Singapore), IIIT Hyderabad and UIUC-ADSC (Singapore). His research focuses on Human-centered computing, especially on modeling non-verbal behavioral cues for interactive analytics. He is an IEEE Senior Member, and an ACM and AAAC member.
\end{IEEEbiography}
\vskip -3\baselineskip plus -1fil
\begin{IEEEbiography}[{\includegraphics[width=1in,height=1.25in,keepaspectratio]{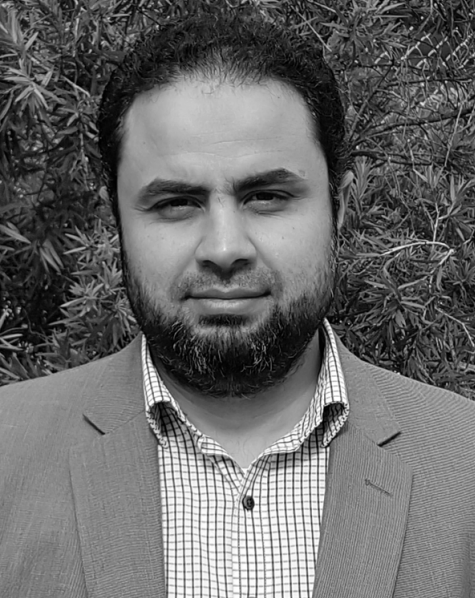}}]{Ibrahim Radwan} received the Ph.D. degree in computer science from the University of Canberra in 2015. From 2014 to 2016, he was a researcher in a leading automotive industry warehouse, Research Fellow with The Australian National University and currently an assistant professor at the University of Canberra. His research includes computer vision, machine learning, robotics, and artificial intelligence.
\end{IEEEbiography}
\vskip -3\baselineskip plus -1fil
\begin{IEEEbiography}[{\includegraphics[width=1in,height=1.25in,keepaspectratio]{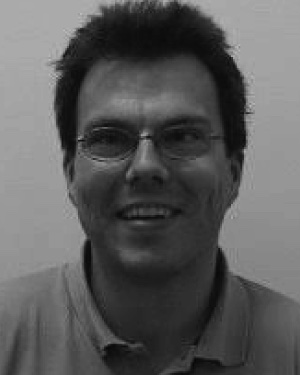}}]{Roland Goecke} received his Ph.D. degree in computer science from The Australian National University, Canberra, Australia, in 2004. He is Professor of Affective Computing with the University of Canberra, where he is serves as Director of the Human-Centred Technology Research Centre. His research interests include affective computing, pattern recognition, computer vision, human–computer interaction and multimodal signal processing. He is a senior member of the IEEE, and an ACM and AAAC member.
\end{IEEEbiography}

\end{document}